\documentclass[twocolumn,showpacs,preprintnumbers,superscriptaddress,amsmath,amssymb]{revtex4}
\usepackage{epsfig}
\usepackage{subfigure}
\usepackage{amsmath}
\usepackage{amssymb}
\usepackage{graphicx}
\usepackage{dcolumn}
\usepackage{bm}
\input epsf

\begin{document}

\title{The onset of synchronization in large networks of coupled oscillators}

\author{Juan G. Restrepo}
\email{juanga@math.umd.edu}
\affiliation{
Institute for Research in Electronics and Applied Physics, 
University of Maryland, College
Park, Maryland 20742
}
\affiliation{
Department of Mathematics,
University of Maryland, College Park, Maryland 20742
}

\author{Edward Ott}
\affiliation{
Institute for Research in Electronics and Applied Physics, 
University of Maryland, College
Park, Maryland 20742
}
\affiliation{
Department of Physics and Department of Electrical and Computer Engineering,
University of Maryland, College Park, Maryland 20742
}

\author{Brian R. Hunt}
\affiliation{
Department of Mathematics,
University of Maryland, College Park, Maryland 20742
}
\affiliation{
Institute for Physical Science and Technology,
University of Maryland, College Park, Maryland 20742
}

\date{\today}

\begin{abstract}
We study the transition from incoherence to coherence in large networks of coupled phase oscillators. 
We present various approximations that describe the behavior of an appropriately defined order parameter
past the transition, and generalize recent results for the critical coupling strength.
We find that, under appropriate conditions,
the coupling strength at which the transition occurs is determined by the largest eigenvalue of the adjacency matrix. 
We show how, with an additional assumption, a mean field approximation recently proposed 
is recovered from our results. We test our theory with numerical simulations, and find that it describes the transition when
our assumptions are satisfied. 
We find that our theory describes the transition well in situations in which the mean field approximation fails.
We study the finite size effects caused by nodes with small degree and find that they cause the critical coupling strength 
to increase. 
\end{abstract}


\maketitle

\section{Introduction}

In recent years, the importance of networks in different fields 
has become increasingly clear \cite{newman1, barabasi1,barabasi2}. It has been observed 
that many real world networks possess topologies which introduce important effects on the processes taking place
on them. One of the most interesting and important of these processes is the synchronization of coupled dynamical 
systems.
Synchronization is found in fields ranging from physics to biology \cite{pikovsky,mosekilde}, 
and in many cases involves a large network of dynamical 
systems. The structure of this network plays a crucial role in determining the synchronization of the coupled elements.

Kuramoto \cite{kuramoto} proposed and exactly solved a model for the synchronization of all-to-all uniformly 
coupled phase oscillators. His model and solution
have become a guide as to how the coupling strength and the properties of the oscillators (e.g., their natural frequencies) 
might affect their synchronization, and generalizations of this basic model have been studied \cite{strogatz,pikovsky}.
Some attempts to study the Kuramoto model with networks different from the all-to-all network have been made \cite{jadbabaie}.
Networks in which the interaction strength depends on a distance have been studied, and it has
been numerically found that a transition from incoherent to coherent behavior occurs at a critical value of the coupling
strength  \cite{bahiana}. 
The Kuramoto model in networks without global coupling has recently started to receive attention. It was numerically observed \cite{moreno}
that a transition is also present in scale free networks.  
Very recently, a mean field theory to determine the 
transition to synchronization in more general networks has been proposed \cite{ichinomiya,lee}.
The mean field theory result is that the critical coupling strength $k_{mf}$ is determined by the Kuramoto value, $k_0$, 
rescaled appropriately by the first two moments
of the degree distribution of the nodes in the network: $k_{mf} = k_0 \langle d \rangle/\langle d^2 \rangle$,
where
\begin{equation}\label{defi}
\langle d^q \rangle = \frac{1}{N}\sum_{n=1}^N d_n^q,
\end{equation}
the degree $d_n$ of node $n$ is the number of connections between node $n$ and other nodes of the network, and $N$ is the number
of nodes in the network. 
 
In this paper we go beyond the mean field approximation, obtaining a better estimate of the critical coupling strength. We also 
describe
the behavior of a suitably defined order parameter past the transition. 
We  show how our results reduce to those of the mean field theory when an additional assumption is introduced, 
and present examples in different regimes. We find that in some
regimes the mean field approximation does not provide an adequate description of the transition,
whereas our more general estimate does. We also show how our results explain
observations for networks with distance dependent interaction strength.
We study finite size effects caused mainly by nodes of small degree, and find that the transition point is shifted
to larger values of the coupling strength when these effects are taken into account.

This paper is organized as follows. In Section~\ref{self} we present our theory and discuss the mean field approach. 
In Section~\ref{examples}, we present numerical examples for different situations and test the different approximations.
In Section~\ref{nonuniform} we discuss the case of networks with nonuniform coupling strength.
In Section~\ref{linear}, we present a linear analysis of the problem. In Section~\ref{fluctuations} we consider
finite size effects caused primarily by nodes with a small number of connections.
Finally, we conclude in Section~\ref{discussion}. Some calculations were relegated to Appendices \ref{appendixa},
\ref{appendixa2}, and \ref{appendixb}.

\section{Self consistent analysis}\label{self}

As shown by Kuramoto \cite{kuramoto}, the dynamics of weakly coupled, nearly identical limit cycle oscillators 
can, under certain conditions, be approximated by an equation for the phases $\theta_n$ of the form 
\begin{equation}\label{eq:coupled0}
\dot{\theta}_n = \omega_n + \sum_{m=1}^{N} \Omega_{nm}(\theta_{m} - \theta_{n}),
\end{equation}
where $\omega_n$ is the natural frequency of the oscillator $n$, $N$ is the total number of oscillators
and $\Omega_{nm}$ is a periodic function depending on the original equations of motion.
The all-to-all Kuramoto model assumes that $\Omega_{nm}(\theta_m - \theta_n) = (k/N) \sin(\theta_m - \theta_n)$,
where $k$ represents an overall coupling strength.
In order to incorporate the presence of a heterogeneous network, we assume that 
$\Omega_{nm}(\theta_m - \theta_n) =  k A_{nm} \sin(\theta_m - \theta_n)$, where $A_{nm}$ are the elements of a $N \times N$ 
adjacency matrix $A$ determining the connectivity of the network.
Therefore, we study the system
\begin{equation}\label{eq:coupled}
\dot{\theta}_n = \omega_n + k \sum_{m=1}^{N} A_{nm}\sin(\theta_{m} - \theta_{n}).
\end{equation}

For specificity, we will primarily consider the case where the $A_{nm}$ are either $0$ (nodes $n$ and $m$ are not connected)
or $1$ (nodes $n$ and $m$ are connected, and all connections have equal strength).
We assume that the network is undirected, so that $A_{nm} = A_{mn}$.
We assume also that, for each $n$, the corresponding $\omega_n$ is independently chosen from a known
oscillation frequency probability distribution $g(\omega)$.    
We assume that $g(\omega)$ is symmetric about a single local maximum (cf. Sec~\ref{linear}), which without loss of generality
we can take to be at $\omega = 0$. (If the mean frequency is $\omega_0 \neq 0$, we make the change of
coordinates that shifts each $\omega_n$ by $\omega_0$ and each $\theta_n$ by $\omega_0 t$.) In this case, synchronization
will occur at frequency $0$, i.e., $\theta_n$ will remain approximately constant for synchronized nodes.

We define a positive real valued local order parameter $r_{n}$ by 
\begin{equation}\label{eq:rn}
r_n e^{i \psi_n} \equiv \sum_{m=1}^{N} A_{nm}\langle e^{i \theta_m}\rangle_t,
\end{equation}
where $\langle\dots\rangle_t$ denotes a time average.
In terms of $r_n$, Eq.~(\ref{eq:coupled}) can be rewritten as
\begin{equation}\label{eq:coupled2}
\dot{\theta}_n = \omega_n - k r_{n} \sin(\theta_n - \psi_n) - k h_n(t),
\end{equation}
where the term $h_n(t)$ takes into account time fluctuations and is given by
$h_n = Im \{ e^{-i\theta_n}\sum_m A_{nm}\left( \langle e^{i\theta_m}\rangle_t - e^{i\theta_m}\right)\} $,
where $Im$ stands for the imaginary part.
Since we regard $h_n$ as a sum of 
$d_n$ approximately uncorrelated terms 
(where $d_n$ is the degree of node $n$ given by $d_n = \sum_m A_{nm}$), we expect $h_n$ to be of order $ \sqrt{d_n}$.
Substantially above the transition, due to the synchronization of the phases, 
the quantity $r_n \approx  \sum_m A_{nm} \langle e^{i\theta_m}\rangle_t$
is $\mathcal{O}(d_n)$. Thus, if we assume that $d_{n} \gg 1$, 
substantially above the transition the term $ h_n$ can be neglected with respect to $r_n$. However, just
above the transition to coherence,
the number of oscillators that are phase locked is small (see below), and so the term $r_n$ is also small. We need the number of 
locked oscillators to be large enough so that we can neglect $h_n$, but, in cases where we use perturbative methods,
we also require that the number of locked oscillators be
small enough that the perturbative methods are still valid. We therefore do not expect 
the perturbative methods to agree perfectly just at the transition point.
[Indeed in the classical Kuramoto (all-to-all) model a similar reservation holds for finite networks, as there are
 $\mathcal{O}(N^{-1/2})$ fluctuations of $k\sum_{m=1}^N e^{i\theta_m}$ for $k$ below its critical transition value.]
In Sec.~\ref{fluctuations} we will investigate the effects of the time fluctuating term $h_n$ in Eq.~(\ref{eq:coupled2}), 
but, for now, we neglect it.

With $h_n$ neglected in Eq.~(\ref{eq:coupled2}), 
oscillators with $\left|\omega_n\right| \leq k r_n$ become locked,
i.e., for these oscillators $\theta_n$ settles at a value for which
\begin{equation}\label{eq:locked}
\sin(\theta_{n}-\psi_n) = \omega_n/(k r_n).
\end{equation}
(In general there are two such $\theta_n$; the one closest to $\psi_n$ is stable.) Then
\begin{equation}\label{eq:beta}
r_{n} = \sum_{m=1}^{N} A_{nm} \langle e^{i(\theta_m - \psi_n)}\rangle_t
\end{equation}
\begin{eqnarray}
 =  \sum_{\left|\omega_{m}\right| \leq k r_{m}} A_{nm}  e^{i(\theta_m - \psi_n)}\nonumber\\
+   \sum_{\left|\omega_{m}\right| > k r_{m}} A_{nm} \langle e^{i(\theta_m - \psi_n)}\rangle_t.\nonumber
\end{eqnarray}
In order to proceed further, we will introduce the following assumption:

\vspace{5mm}\noindent{\bf Assumption} $\bigstar$ {\it We assume the existence of solutions $r_n$, $\psi_n$ 
that are statistically independent of $\omega_{n}$.}
\vspace{5mm} 

\noindent This is a nontrivial assumption; however, it is reasonable if most of node $n$'s neighbors have reasonably
large degree, so that they are not strongly affected by the value of $\omega_n$. And, as we show below, such a solution can 
be found in a self consistent manner. 
Using a milder version of Assumption~$\bigstar$, we show in Appendix A that 
the sum over the unlocked oscillators in Eq.~(\ref{eq:beta}) can be neglected. 
Therefore, only the locked oscillators remain in the sum, and we get from Eq.~(\ref{eq:beta}) 
using Eq.~(\ref{eq:locked}), since $r_n$ is by definition real,
\begin{eqnarray}\label{eq:betaprime}
r_n = Re \{ \sum_{\left|\omega_{m}\right| \leq k r_{m}} A_{nm} e^{i(\theta_m - \psi_m)}e^{i(\psi_m - \psi_n)}   \}\\
= \sum_{\left|\omega_{m}\right| \leq k r_{m} }\nonumber 
 A_{nm} \cos(\psi_m - \psi_n)\sqrt{1 - \left(\frac{\omega_{m}}{k r_{m}}\right)^2}\\
- \sum_{\left|\omega_{m}\right| \leq k r_{m} } 
 A_{nm} \sin(\psi_m - \psi_n)\left(\frac{\omega_{m}}{k r_{m}}\right),\nonumber
\end{eqnarray}
where $Re$ represents the real part.
For the imaginary part of Eq.~(\ref{eq:beta}), we get
\begin{eqnarray}\label{eq:imaginary}
0 = \sum_{\left|\omega_{m}\right| \leq k r_{m}  } 
 A_{nm} \cos(\psi_m - \psi_n) \left(\frac{\omega_{m}}{k r_{m}}\right)\\
+ \sum_{\left|\omega_{m}\right| \leq k r_{m}  } 
 A_{nm} \sin(\psi_m - \psi_n)\sqrt{1-\left(\frac{\omega_{m}}{k r_{m}}\right)^2}.\nonumber
\end{eqnarray}

Using Assumption~$\bigstar$, the contribution
of the last term in the real part equation (\ref{eq:betaprime}) can be neglected because 
of the symmetry of $g(\omega)$ about $0$. We thus obtain the approximation
\begin{equation}\label{eq:cosfi}
r_n = \sum_{\left|\omega_{m}\right| \leq k r_{m} } 
 A_{nm} \cos(\psi_m - \psi_n)\sqrt{1 - \left(\frac{\omega_{m}}{k r_{m}}\right)^2}.
\end{equation}

Since we are interested in the transition to coherence, we look for the solution of Eq.~(\ref{eq:cosfi})
that yields the smallest critical coupling $k$. The smallest critical coupling is obtained when the cosine
in Eq.~(\ref{eq:cosfi}) is $1$. (Note that both the number of terms in the sum and their size decreases as $k$ decreases. 
Hence, a smaller $k$ corresponds to a larger value of the cosine.)
We therefore will look for solutions for which $\psi_n-\psi_m = 0$, i.e., $\psi_n$ does not depend on $n$, 
and without loss of generality,
we will take $\psi_n \equiv 0$.
Note that this is a consistent condition in the 
sense that the imaginary part equation (\ref{eq:imaginary}) is satisfied:  
the first term vanishes in the limit of a
large number of connections per node due to the symmetry around $0$ of $g(\omega)$, 
and the second due to our assumed form that $\psi_n$ does not depend on $n$.

Equation (\ref{eq:cosfi}) then reduces to
\begin{equation}\label{eq:betass}
r_n = \sum_{\left|\omega_{m}\right| \leq k r_{m} } 
 A_{nm} \sqrt{1 - \left(\frac{\omega_{m}}{k r_{m}}\right)^2}.
\end{equation}
If the particular collection of frequencies $\omega_n$ is known, this equation can be solved numerically. We will refer 
to this approximation, based on neglecting the time fluctuations in Eq.~(\ref{eq:coupled2}),
 as the {\it time averaged theory} (TAT). 
We now define an order parameter $r$ by
\begin{equation}\label{eq:orderpara}
r = \frac{\sum_{n=1}^{N} r_{n}}{\sum_{n=1}^{N} d_{n}},
\end{equation}
where $d_n$ is the degree of node $n$ defined by $d_n = \sum_{m=1}^N A_{nm}$.
Note that $r = \sum_{n=1}^N d_n \langle e^{i\theta_n}\rangle_t/\sum_{n=1}^{N} d_{n}$ coincides with the order parameter used in
Refs.~\cite{ichinomiya,lee}.

If the number of connections per node is large, the particular collection of frequencies
of the neighbors of a given node will likely be a faithful sample of the frequency distribution $g(\omega)$. Assuming this
is the case, and using Assumption~$\bigstar$, we approximate the sum in Eq.~(\ref{eq:betass}) as
\begin{equation}\label{eq:betanoprime}
r_{n} = {\sum_{m}} A_{nm} \int_{- k r_{m}}^{k r_{m}}
 g(\omega) \sqrt{1 -\left(\frac{\omega}{k r_{m}}\right)^2 } d\omega,
\end{equation}
or, introducing $z\equiv \omega/(k r_m)$,
\begin{equation}\label{eq:betaint} 
r_{n} = k {\sum_{m}} A_{nm} r_{m} \int_{-1}^{1}
 g(z k r_{m}) \sqrt{1 - z^2 } dz
\end{equation} 
This equation is one of our main results. It is analogous to Eq.~(13) in Ref.~\cite{ichinomiya} and 
Eq.~(6) in Ref.~\cite{lee}, but, as opposed to including
only information of the degree distribution of the network, it depends on the adjacency matrix, which 
completely describes the topology of the network. Equation~(\ref{eq:betaint}) determines 
implicitly the order parameter $r$ as a function of 
the network $A_{nm}$, the frequency distribution $g(\omega)$, and the coupling constant $k$.
We will refer to this approximation 
as the {\it frequency distribution approximation} (FDA). 
As with the TAT approximation (\ref{eq:betass}), nonlinear matrix equation (\ref{eq:betaint}) 
can be solved numerically and the order parameter $r$ computed from $r_n$ using Eq.~(\ref{eq:orderpara}).

We will now study the implications of Eq.~(\ref{eq:betaint}) by using approximation schemes in different 
regimes in order to obtain explicit expressions for the order parameter and the critical coupling strength.

\subsection{Perturbation Theory (PT)}\label{perturbative}

From the discussion above, coherent behavior is characterized by a nonzero value of $r_{n}$. 
We determine the critical value of $k$ by letting $r_{n} \to 0^+$.
The first order approximation $g(zkr_m) \approx g(0)$ in Eq.~(\ref{eq:betaint}) produces 
\begin{equation}\label{eq:firstor}
r_{n}^{(0)} = \frac{k}{k_0} {\sum_{m}} A_{nm} r_{m}^{(0)},
\end{equation}
where $k_{0} \equiv 2/(\pi g(0))$.
Since we are interested in the transition to coherence, the smallest $k$ satisfying Eq.~(\ref{eq:firstor}) is
of interest.
We thus identify the critical transition value of 
$k_0/k$ with the largest eigenvalue $\lambda$ of the adjacency matrix $A$, obtaining 
\begin{equation}\label{eq:kc}
k_{c} = \frac{k_0}{\lambda}.
\end{equation}
(In the case $A_{nm} \equiv 1$ of all-to-all coupling, $\lambda = N-1$.)
Also $r_{m}^{(0)}$ is proportional to the $m$th component of the eigenvector $u = [u_1,u_2,\dots,u_N]^T$ 
associated with this eigenvalue. Note that this 
is consistent with Assumption~$\bigstar$, since $r_{n}$ depends only on network properties
(i.e., the matrix $A$) and is thus independent of $\omega_{n}$. 
Equation (\ref{eq:kc}) is one of our main results. It determines when the transition
to coherence occurs in terms of the largest eigenvalue $\lambda$ of the adjacency matrix $A$.

In order to assess how the order parameter $r$ given by Eq.~(\ref{eq:orderpara}) grows as $k$ grows from $k_c$,
we must take into account that $g(zkr_m)$ in Eq.~(\ref{eq:betaint}) is not constant.
For $k r_{n}$ small (see the discussion at the end of Sec.~\ref{mftsec}),
the second order approximation yields
\begin{align}
r_n = k {\sum_{m}} A_{nm} r_{m} \hspace{2cm}\\
\times \int_{-1}^{1}
 \left(g(0) + \frac{1}{2}g''(0)(z k r_m)^2\right) \sqrt{1 - z^2 } dz.\nonumber
\end{align}
Defining $\alpha \equiv -\pi g''(0)k_{0}/16$, we get
\begin{equation}\label{eq:pertu}
r_n = \frac{k}{k_{c}\lambda} \sum_{m} A_{nm} \left(r_m
 - \alpha k^2 r_m^3\right).
\end{equation}
We consider perturbations from the first order critical values as follows:
\begin{eqnarray}
r_n = r_n^{(0)} + \delta r_n,
\end{eqnarray}
where $\delta r_n \ll r_n^{(0)} \ll 1$ as $k \to k_c$.
Inserting this into Eq.~(\ref{eq:pertu}), and canceling terms of order $r_n^{(0)}$, 
the leading order terms remaining are
\begin{eqnarray}\label{eq:fea}
\delta r_n = \frac{k}{k_{c}\lambda} \sum_{m} A_{nm}\delta r_m -
\frac{\alpha k^3}{k_{c}\lambda}\sum_{m} A_{nm} (r_m^{(0)})^3 \\
+ \frac{k - k_{c}}{k_c \lambda} \sum_{m} A_{nm} r_m^{(0)}\nonumber.
\end{eqnarray}
In order for Eq.~(\ref{eq:fea}) to have a solution for $\delta r_n$, it must satisfy a solubility
condition. This condition can be obtained by
multiplying by $r_n^{(0)}$, summing over $n$, using Eq.~(\ref{eq:firstor})
and the assumed symmetry $A_{nm} = A_{mn}$, to obtain
\begin{equation}
\frac{\sum_{m} (r_m^{(0)})^4}{\sum_{m} (r_m^{(0)})^2} = \frac{k - k_c}{\alpha k^3}.
\end{equation}
In terms of $u$, the normalized eigenvector of $A$ associated with the eigenvalue $\lambda$, 
the square of the order parameter $r$ can be expressed as
\begin{equation}\label{eq:secondpt}
r^2 = \left(\frac{\eta_{1}}{\alpha k_{0}^2}\right)
\left(\frac{k}{k_{c}} - 1\right)
\left(\frac{k}{k_{c}}\right)^{-3}
\end{equation}
for $k/k_c >1$, where
\begin{equation}\label{eq:eta1}
\eta_1 \equiv \frac{\langle u\rangle^2 \lambda^2}{ N \langle d\rangle^2 \langle u^4\rangle}.
\end{equation}

Eqs.~(\ref{eq:secondpt}) and (\ref{eq:eta1}) describe the behavior of the order parameter near the transition in terms 
of $\lambda$ and its associated eigenvector. 
We will refer to them as the {\it perturbation theory} (PT).

The presence of the term $\langle u^4\rangle$ in Eq.~(\ref{eq:eta1}) suggests that the expansion of $g$ to second
order might fail when there are a few components of the eigenvector $u$ that are much larger than the rest. 
This occurs when the degree distribution is highly heterogeneous. We formulate more precisely this constraint 
in the discussion at the end of Sec.~\ref{mftsec}.

\subsection{Mean field theory (MF)}\label{mftsec}

In this section we describe an approximation that works in some regimes and has the advantage of greater
analytical tractability. In this section we also recover some of the results in Refs.~\cite{ichinomiya,lee}.
Here we assume that $r_n$ is proportional to $d_n$, $r_n \propto d_{n}$. 
The assumption consists in treating the average
\begin{equation}\label{eq:meanfield}
\frac{r_n}{d_n} = \frac{1}{d_{n}} \left|\sum_{m=1}^{N} A_{nm} \langle e^{i\theta_{m}}\rangle_t \right|,
\end{equation}
which depends on $n$, as if it were a constant independent of $n$. Following Refs.~\cite{ichinomiya,lee},
we call this the {\it mean field} (MF) approximation. 
It is also equivalent, near the transition, to
assuming that the eigenvector associated with the largest eigenvalue $\lambda$ satisfies $u_n\propto d_n$. 
We will discuss later the range of validity of this assumption.
Note that this form for $r_n$ is 
again consistent with our Assumption~$\bigstar$ that $r_n$ is independent of $\omega_{n}$.
The ratio $r_n/d_n$ coincides under this approximation with the order parameter $r$ defined in Eq.~(\ref{eq:orderpara}).

Summing over $n$ and substituting $r_n = r d_{n}$ in Eq.~(\ref{eq:betaint}), we obtain  
\begin{equation}\label{eq:betasumed}
\sum_{m = 1}^N d_{m} = k \sum_{m = 1}^N d_{m}^2 \int_{-1}^{1} g(z k r d_{m}) \sqrt{1 - z^2 } dz,
\end{equation}
which coincides with Eq.~(13) in Ref.~\cite{ichinomiya}. 
As we approach the transition from above, $r \to 0^+$, the first order approximation is 
$g(z k r d_{m}) \approx g(0)$, from which we obtain
\begin{equation}\label{eq:firstorder}
k \equiv k_{mf} = k_{0} \frac{\langle d \rangle}{\langle d^2 \rangle},
\end{equation}
the main result of Ref.~\cite{ichinomiya}.

In the limit $N\to \infty$, we can replace $\langle d^q \rangle$ as defined by Eq.~(\ref{defi}) by 
\begin{equation}\label{x}
\langle d^q \rangle_{\infty} = \int d^q p(d) dd,
\end{equation}
where $p(d)$ is the probability distribution function for the degree. 
Note that from Eq.~(\ref{defi}), $\langle d^q \rangle$ is always well-defined for finite $N$,
but that Eq.~(\ref{x}) indicates that $\langle d^q\rangle_{\infty}$ diverges for
power law degree distributions $p(d) \propto d^{-\gamma}$ if $\gamma \leq q + 1$. We also note that 
many real networks have approximate power law $p(d)$ with $\gamma < 3$ (see Ref.~\cite{newman1}). 
On the basis that $\langle d^2 \rangle_{\infty}/ \langle d \rangle_{\infty}= \infty$ for $2 \leq \gamma \leq 3$, Ichinomiya \cite{ichinomiya}
notes that from Eq.~(\ref{eq:firstorder}) $k_{mf}\to 0$ as $N \to \infty$; i.e., predicts that in the limit 
$N\to \infty$ there is no threshold for coherent oscillations when $2\leq \gamma \leq 3$.  
As will become evident,
our numerical experiments, although for $N \gg 1$, are often not well-approximated by the $N\to \infty$ limit,
in particular for $\gamma < 3$.

The mean field approximation can be pushed further to second order by
expanding $g(z k r d_{m}) \approx g(0) + \frac{1}{2} g''(0)(z k r d_{m})^2 $ in Eq.~(\ref{eq:betasumed}),
obtaining, provided $k r d_{m}$ is small,
\begin{equation}\label{eq:secondorder}
1 = \frac{k}{k_{mf}} + k^3 r^2 \frac{\pi}{16} g''(0) \frac{\sum_{m = 1}^N d_{m}^4}{\sum_{m = 1}^N d_{m}},
\end{equation}
so that
\begin{equation}\label{eq:secondmf}
r^2 =  \left(\frac{\eta_{2}}{\alpha k_{0}^2}\right) 
\left(\frac{k}{k_{mf}} - 1\right)\left( \frac{k}{k_{mf}}\right)^{-3}
\end{equation}
for $k/k_c > 1$, where
\begin{equation}\label{eq:eta2}
\eta_{2} \equiv \frac{\langle d^2\rangle^3}{\langle d^4\rangle \langle d \rangle^2}.
\end{equation}
In expanding $g$ to second order, it was assumed that $kd_m$ is small. The term $\langle d^4 \rangle$
in Eq.~(\ref{eq:eta2}) suggests that the conditions under which the expansion of $g$ is appropriate are those 
under which $\langle d^4 \rangle_{\infty}$ is finite. In fact, Lee shows \cite{lee} that for a power law distribution of the 
degrees, $p(d) \propto d^{-\gamma}$, the above expansion is appropriate for $\gamma > 5$. For $3 \leq \gamma \leq 5$,
he obtains in the limit $N\to\infty$ that $r$ scales near the transition as 
$r \propto \left(\frac{k}{k_{mf}} - 1\right)^{1/(\gamma-3)}$. A similar situation occurs in the perturbation 
theory [Eqs.~(\ref{eq:secondpt}) and (\ref{eq:eta1})], which was also based on expanding $g$ to second order. According to 
the previous discussion, we will only use the expression for $r$ obtained from the perturbation theory for situations in 
which $\langle d^4\rangle_{\infty}$ is finite. The critical coupling strength in Eq.~(\ref{eq:kc}), on the other hand, 
does not have this restriction.

The expressions in Eqs.~(\ref{eq:eta1}) and (\ref{eq:eta2}) can be shown to coincide under the approximation $u_n \propto d_n$.
The treatment in Section \ref{perturbative} does not assume that $r_{n}/d_n$ is independent of $n$, 
and we will show in Section \ref{examples} that there are significant cases
where it gives better results for the critical coupling strength than the mean field approximation.

\subsection{Summary of approximations and range of validity}\label{rangeofvalidity}

In the previous sections, we developed different approximations to find the critical coupling constant and the 
behavior of the order parameter past the transition. Here we summarize the different approximations and the
assumptions used in obtaining them. All the approximations mentioned above 
assume that the number of connections per node is very large.
This allowed us, among other things, to neglect the time fluctuating term $h_n(t)$ in Eq.~(\ref{eq:coupled2}). 
We will discuss the effect of this
term in Section \ref{fluctuations}.

\begin{figure}[h]
\begin{center}
\epsfig{file = 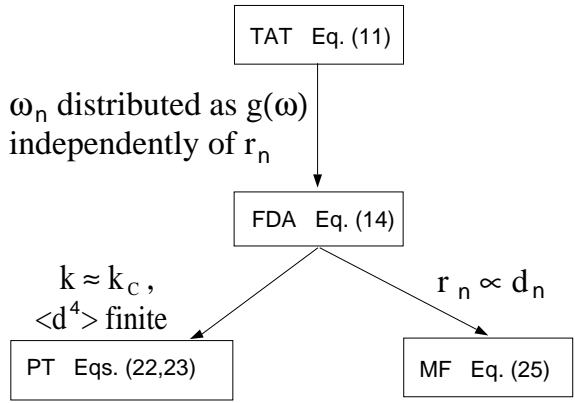, clip =  ,width=0.9\linewidth}
\caption{Different approximations and the assumptions leading to them. See text for details.} 
\label{fig:diagram}
\end{center}
\end{figure}

The most fundamental approximation is given by Eq.~(\ref{eq:betass}). 
This equation can be solved numerically if the
frequency of each oscillator and the adjacency matrix is known. This is the time averaged theory (TAT). 
Assuming that the local mean field $r_n$ is statistically independent of the frequency $\omega_n$, the
frequency distribution approximation (FDA) given by Eq.~(\ref{eq:betaint}) is obtained.
This equation can also be solved numerically, but only knowledge of the probability distribution for
the frequencies and the adjacency matrix is required.
Obtained by expanding the FDA approximation near the transition point, 
the perturbation theory (PT) describes the behavior of the order parameter in terms 
of the largest eigenvalue of the adjacency matrix and its associated eigenvector
in networks where the degree distribution is relatively homogeneous, 
more precisely when $\langle d^4\rangle_{\infty}$ is finite.
Taking $r_n$ in the FDA approximation to be proportional to the degree, $r_n \propto d_n$, 
leads to the mean field theory (MF).
Table~\ref{abbreviations} summarizes the different approximations, their abbreviations and their corresponding equations. 
The diagram in Fig.~\ref{fig:diagram} indicates the assumptions leading to each approximation. 

\begin{table}
\begin{tabular}{|c|c|c|}
\hline
Approximation& Abbreviation & Equation\\
\hline
Time averaged theory & TAT & (\ref{eq:betass}) \\
\hline
Frequency distribution  &FDA & (\ref{eq:betaint}) \\
approximation& &\\
\hline
Perturbation theory &PT &(\ref{eq:secondpt},\ref{eq:eta1})\\
\hline
Mean field theory & MF & (\ref{eq:betasumed})\\
\hline
\end{tabular}
\caption{Approximations considered, their abbreviation, and their corresponding equations.}
\label{abbreviations}
\end{table}

The mean field theory requires only knowledge of the frequency distribution and the degree distribution of the network, and thus 
it requires less information than the other approximations. However, it can produce misleading results if not used carefully.
The mean field approximation has the added assumption that the 
eigenvector $u$ of $A$ associated with the largest
eigenvalue $\lambda$ satisfies $u_n \propto d_n$ (since, close to the transition, $r_n \approx u_n$). 
While correlations might exist \cite{goh}, 
these two quantities are in general not proportional. 
Further, the mean field approximation implies that $\lambda \approx \langle d^2 \rangle/\langle d \rangle$,
a result that, although a good approximation in some cases, is not always true.
Asymptotic forms for
the largest eigenvalue in random networks with given degree distributions are discussed and a sufficient condition
for $\lambda \approx \langle d^2 \rangle/\langle d \rangle$ to be valid is presented in \cite{chung} as follows.
Let $d_{max}$ be the maximum expected degree of the network. If $\langle d^2 \rangle/\langle d \rangle > \sqrt{d_{max}} \log N$,
then $\lambda \approx \langle d^2 \rangle/\langle d \rangle$ almost surely as $N\to \infty$.
We note also that, if the degree distribution is tightly distributed around its mean, so that
$\sqrt{\langle d^2} \rangle \sim  \langle d \rangle \sim d_{max} \gg (\log N)^2$, the condition for
the validity of $\lambda \approx \langle d^2 \rangle/\langle d \rangle$ is satisfied. 
If instead $\sqrt{d_{max}} > (\langle d^2 \rangle/\langle d \rangle) (\log N)^2$, then almost surely  
the largest eigenvalue is $\lambda \approx \sqrt{d_{max}}$ as $N\to \infty$ \cite{chung}.
We will show that, indeed, to the extent that the approximation $\lambda \approx \langle d^2 \rangle/\langle d \rangle$
does not hold, the results from the numerical 
simulation of Eq.~(\ref{eq:coupled}) agree with the critical coupling strength as determined by the 
eigenvalue of the adjacency matrix, rather than by the quantity
$\langle d^2 \rangle/\langle d \rangle$.

\begin{figure}[h]
\begin{center}
\epsfig{file = 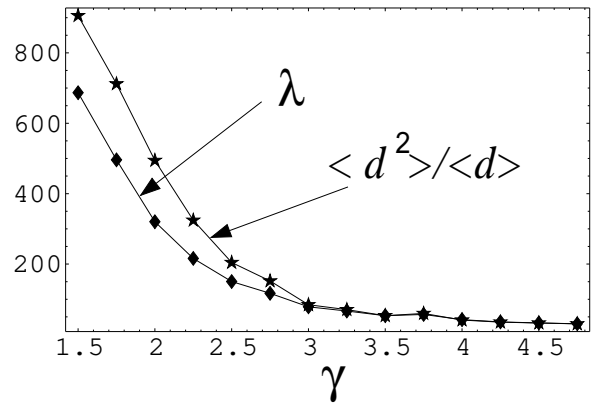, clip =  ,width=0.9\linewidth}
\caption{
Largest eigenvalue $\lambda$ (diamonds) and $\langle d^2 \rangle/\langle d \rangle$ (stars) as a function of $\gamma$ for
$N = 5000$ and $d_0 = 20$.
} 
\label{fig:fignume}
\end{center}
\end{figure}

The asymptotic regimes described in \cite{chung} are not available with the relatively small networks ($N \sim 5000$) we
are restricted to study due to limited computational resources (see the end of Appendix \ref{appendixa2}). 
Also, finite but large networks are also interesting
from an applied point of view. Thus, we numerically compare both approximations
in order to illustrate the possible discrepancies between them in particular cases.
Figure \ref{fig:fignume} was obtained using (for each $\gamma$) a single random realization of a network where the degrees $d_n$ are 
drawn from a power law degree distribution
with power law exponent $\gamma$ (with $d_n \geq d_0 = 20$) and with $N = 5000$ nodes 
(see Sec.~\ref{examples} for details on how the networks are generated).
We plot $\langle d^2 \rangle/\langle d \rangle$ and $\lambda$
as a function of $\gamma$.
For the parameters used in the plot, $\langle d^2 \rangle/\langle d \rangle$ coincides with the largest eigenvalue $\lambda$
for values of $\gamma$ greater than $3$. This suggests that the mean field result for the critical coupling strength $k_{mf}$
is valid for $N = 5000$ and $\gamma > 3$. This is consistent with our numerical experiments in Sec.~\ref{examples}. 
We show in Appendix \ref{appendixa2}, however, that for $\gamma >3$ the mean field approximation $\langle d^2 \rangle/\langle d \rangle$ 
underestimates $\lambda$ for sufficiently large $N$ (too large for us to simulate). In fact, as $N\to \infty$, $\lambda$ 
diverges while $\langle d^2 \rangle/\langle d \rangle$ remains finite in the case $\gamma >3$. 
Thus, the critical coupling constant obtained from
our theory approaches zero as $N\to \infty$, while the one obtained from the mean field theory remains constant. This
suggests that the few nodes with high degree are able, for large enough $N$, to synchronize the network, and that 
these nodes are not taken into account by the mean field theory.

For $\gamma < 3$, we observe from Fig.~\ref{fig:fignume} that $\lambda$ is less than 
$\langle d^2 \rangle/\langle d \rangle$ when $N = 5000$. 
Thus, in this range, the mean field theory predicts a transition for a coupling constant that is 
{\it smaller} than that predicted by the perturbative approach. In the next section we will show, 
for a numerical example in this regime, 
that the transition occurs for a larger coupling than that predicted by the mean field theory.

\section{Examples}\label{examples}

In order to test the results in Sec.~\ref{self}, 
we choose a distribution for the natural frequencies given by $g(\omega) = (3/4) (1-\omega^2)$ for
$-1 < \omega < 1$ and $g(\omega) = 0$ otherwise. In order to generate the network, we specify a degree distribution and 
we use the ``configuration'' model (e.g., Sec.~4.2.1 of Ref.~\cite{newman1} and references therein) to generate a random network
realization with the specified degree distribution: (i) we first generate a {\it degree sequence} by assigning a degree $d_n$
to each node $n$ according to the given distribution; (ii) imagining that each node $n$ is given $d_n$ spokes sticking 
out of it, we choose pairs of spoke ends at random, and connect them.

We consider a fixed number of nodes, $N = 2000$, and the following networks with uniform coupling strength (i.e., $A_{nm} = 1$ or $0$) 
(i) the degrees are uniformly distributed between $50$ and $149$, and 
(ii) the probability of having a degree $d$ is given by $p(d) \propto d^{-\gamma}$ 
if $50 \leq d \leq 2000$ and $p(d) = 0$ otherwise, where $\gamma$ is taken to be $2$, $2.5$, $3$ and $4$.
[Our choice $p(d) = 0$ for $d < 50$ insures that there are no nodes of small degree, and 
suggests that our approximation of neglecting the noise-like, fluctuating quantity $h_n$ in Eq.~(\ref{eq:coupled2}) 
is valid. We return to this issue in Section~\ref{fluctuations}.]

The initial conditions for Eq.~(\ref{eq:coupled}) are chosen randomly in the interval $[0,2\pi]$ and Eq.~(\ref{eq:coupled})
is integrated forward in time until a stationary state is reached (stationary state here means stationary in a statistical sense, 
i.e. the solution might be time dependent but its statistical properties remain constant in time).
From the values of $\theta_n(t)$ obtained for a given $k$, the order parameter $r$ is estimated as 
$r \approx \left|\sum_{m=1}^N d_m \langle e^{i\theta_m}\rangle_t/\sum_{m=1}^{N} d_{m}\right|$, where the time average is taken 
after the system reaches the stationary state. 
(Close to the transition, the time needed to reach the stationary state is very long, so that it is difficult to 
estimate the real value of $r$. This problem also exists in the classical Kuramoto all-to-all model.)
The value of $k$ is then increased and the system is allowed to relax to a stationary state,
and the process is repeated for increasing values of $k$.
\begin{figure}[h]
\begin{center}
\epsfig{file = 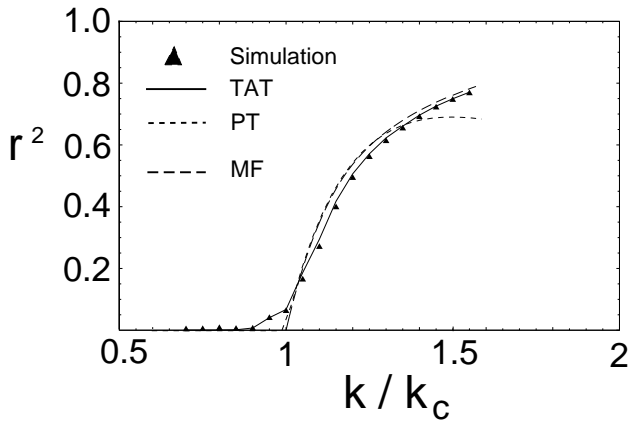, clip =  ,width=1.0\linewidth}
\caption{
Order parameter  $r^2$ obtained from numerical solution of Eq.~(\ref{eq:coupled}) (triangles), time averaged theory (solid line), 
mean field theory (long-dashed line), and perturbation theory (short-dashed line)
as a function of $k/k_c$ for network (i), with the degree of the nodes uniformly distributed 
in $\{50,\dots,149\}$. All curves are obtained using the same single random network realization.} 
\label{fig:figuni}
\end{center}
\end{figure}

In Fig.~\ref{fig:figuni} we show the results for the network with a uniform degree distribution as described above [network (i)].
We plot $r^2$ from numerical solution the full system in Eq.~(\ref{eq:coupled})  (triangles), 
the theoretical prediction from the time averaged theory (solid line),
the prediction from the mean field theory (long-dashed line), and from the perturbation theory (short-dashed line) 
(see Table \ref{abbreviations})
as a function of $k/k_c$, where $k_c$ is given by Eq.~(\ref{eq:kc}).
The frequency distribution approximation agrees with the time averaged theory, so we do not include it in the plot.
In this case, all the theoretical predictions provide good approximations to the observed numerical results.
The time averaged theory reproduces remarkably well the numerical observations. Even the irregular behavior near the transition 
is taken into account by the time averaged theory. The mean field theory is in this case a good approximation, providing a fair
description of the order parameter past the transition. The perturbation theory is valid in this case up to 
$k/k_c \approx 1.3$.

\begin{figure}[t]
\begin{center}
\epsfig{file = 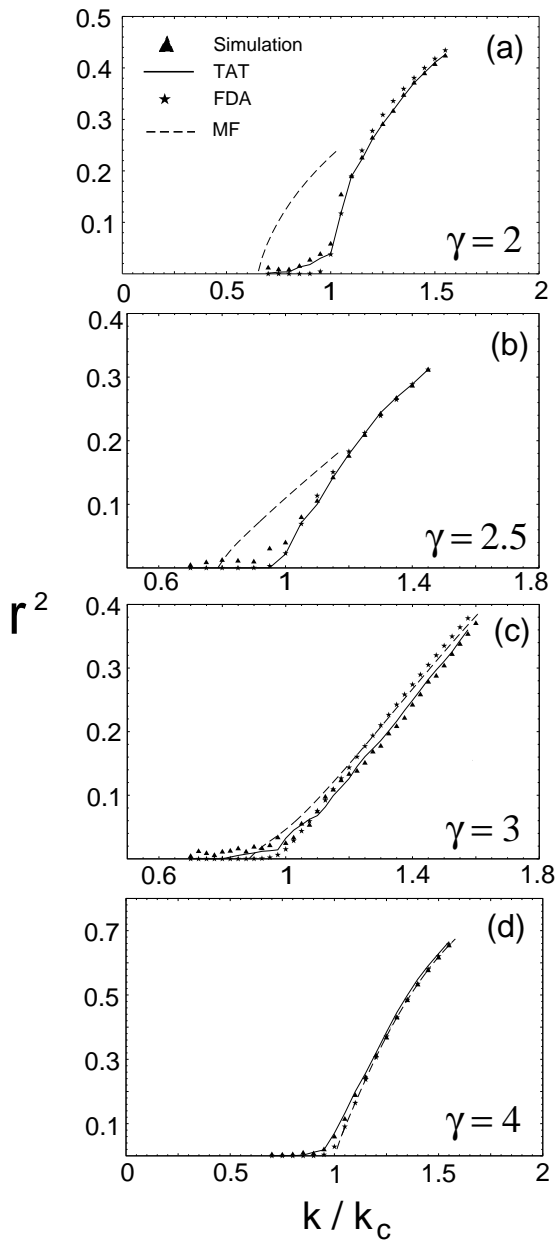, clip =  ,width=0.87\linewidth}
\caption{
Order parameter $r^2$ obtained from numerical solution of Eq.~(\ref{eq:coupled}) (triangles), time averaged theory (solid line), 
frequency distribution approximation (stars), and mean field theory (long-dashed line) 
as a function of $k/k_c$ for degree distributions given by
$p(d) \propto d^{-\gamma}$  if $50 \leq d \leq 2000$ and $p(d) = 0$ otherwise, with (a) $\gamma = 2$, (b) $\gamma = 2.5$, (c) $\gamma = 3$, 
and (d) $\gamma = 4$. All curves in each figure are obtained using the same single random network realization.
} 
\label{figama}
\end{center}
\end{figure}

The results for the networks with power law degree distributions [networks (ii)] are shown in 
Figs.~\ref{figama} (a), (b), (c) and (d) for $\gamma = 2$, $2.5$, $3$, and $4$,
respectively. The order parameter $r^2$ from numerical solution of the full system in Eq.~(\ref{eq:coupled}) (triangles), 
the time averaged theory (solid line), 
the frequency distribution approximation (stars), and the mean field theory (long-dashed line) 
are plotted as a function of $k/k_c$. We do not show the perturbation theory since in all these cases $\gamma < 5$ and so
we do not expect the perturbative theory to be valid as $N\to \infty$.

The time averaged theory agrees best with the numerical simulations in all cases. The frequency distribution
approximation also agrees well in all cases; though it predicts a sharper transition than actually occurs. 
The mean field approximation agrees closely with the frequency 
distribution approximation for $\gamma = 4$ and, away from the transition, for $\gamma = 3$. However,
for $\gamma = 2$ and $\gamma = 2.5$, it deviates greatly from the other approximations and from the numerical simulation. 
The critical coupling strengths predicted by the mean field theory and by the perturbation theory are very close for $\gamma = 4$,
but the mean field theory predicts a transition at about $10\%$ smaller coupling for $\gamma = 3$, about $20\%$ smaller for $\gamma = 2.5$,
and about $40\%$ 
smaller for $\gamma = 2$. Since the transition in the numerical simulation is not so well-defined, both 
approximations are reasonable for $\gamma = 3$, but for $\gamma = 2$ and $\gamma = 2.5$ the critical coupling strength predicted by the mean field
approximation is clearly too small.

In the past years, it has been 
discovered that many real world 
networks have degree distributions which are power laws with exponents between $2$ and $3.5$ \cite{newman1,barabasi1,eguiluz}. 
In order to accurately predict the critical coupling strength across this range of exponents, the critical coupling constant
given by $k_c = k_0/\lambda$ determined by the largest eigenvalue of the adjacency matrix should be used. 
The behavior of the order parameter can be estimated using the time averaged theory or the frequency 
distribution approximation.
These two approximations were found to be consistently accurate for the range of exponents and values 
of the coupling constant studied. 
For the value of $N$ used, the mean field theory works well in predicting the critical coupling strength and the behavior of 
the order parameter if one is interested in values of $\gamma$ larger than $3$. 

Tables \ref{transition} and \ref{orderparameter} present the results of comparing the theoretical predictions
with the numerical integration of Eq.~(\ref{eq:coupled}) for different networks. Table \ref{transition} compares
the observed critical coupling strength with the theoretical estimate. If both are close, the entry is ``G'', and otherwise ``NG''.
Table \ref{orderparameter} compares the predicted behavior of the order parameter past the transition with the observed one. 
If the corresponding entry in Table \ref{transition} is ``NG'', no comparison is attempted. The entries are the range of $k/k_c$
over which the corresponding theoretical prediction agrees with the numerical simulation.

\section{Nonuniform coupling strength}\label{nonuniform}

So far, our examples have assumed that the coupling strength is uniform
(i.e., all the entries of the adjacency matrix $A$ have been taken to be $0$ or $1$). However, 
considering that the degree of a node is defined as $d_n \equiv \sum_{m}A_{nm}$,
our results carry through to the more general case of non uniform coupling.
As an example of this situation, we apply our results to the case treated in Refs.~\cite{bahiana} 
of a distance dependent interaction strength.
Assume that the nodes $n$ are equidistantly located on a circle and the matrix elements are given by 
\begin{equation}  
A_{nm} = f(\left|n-m\right|), 
\end{equation} 
where $\left|n-m\right|$ represents distance modulo $N$ (e.g. $\left|1 - N\right| = 1$), $f(0) = 0$, and $f\geq 0$.
Then each row of $A$ has the same sum $\lambda = \sum_{m} A_{nm}$, and $[1,1,\dots,1]^T$ is an eigenvector with 
eigenvalue $\lambda$. 
By the Gershgorin circle theorem \cite{gershgorin} (each eigenvalue $\sigma$ of $A$ satisfies, for some $n$, 
$\left|\sigma - A_{nn}\right| \leq \sum_{m \neq n} \left|A_{nm}\right|$), this is the largest
eigenvalue (since $A_{nn} = 0$), and thus determines the transition to synchrony as described in the previous section. 
This scaling factor has been proposed before, by analogy to spin systems, to determine the 
transition to coherence in the case of a power law 
decaying interaction strength $f(x) = x^{-\gamma}$ \cite{bahiana}.

\begin{table}
\begin{tabular}{|c|c|c|c|c|c|}
\hline
Degree distribution &TAT & FDA  & MF & PT \\
\hline
$p(d)$ uniform in $\{50,\dots,149\}$ &G&G&G&G \\
\hline
$p(d) \propto d^{-\gamma}$, $\gamma = 2$ &G&G&NG&G\\
\hline
$p(d) \propto d^{-\gamma}$, $\gamma = 3$&G&G&G&G\\
\hline
$p(d) \propto d^{-\gamma}$, $\gamma = 4$ &G&G&G&G \\
\hline
\end{tabular}
\caption{Comparison of the predicted critical coupling strength versus the observed one for the different approximations (columns) and 
different networks (rows). If the critical coupling strength is predicted by a given approximation for a certain network, the corresponding
entry is marked ``G''. Otherwise, ``NG'' is entered. }
\label{transition}
\end{table}

\begin{table}
\begin{tabular}{|c|c|c|c|c|  }
\hline
p(d)& TAT & FDA &  MF & PT \\
\hline
$p(d)$ uniform in $\{50,\dots,149\}$ & $0.5+$ & $0.5+$  & $0.5+$&$0.3$\\
\hline
$p(d) \propto d^{-\gamma}$, $\gamma = 2$ &$0.7+$  & $0.7+$ & -  & -    \\
\hline
$p(d) \propto d^{-\gamma}$, $ \gamma = 3$&$0.7+$ &$0.7+$ & $0.7+$   & - \\
\hline
$p(d) \propto d^{-\gamma}$, $\gamma = 4$ & $0.7+$ & $0.7+$ & $0.7+$ & -     \\
\hline 
\end{tabular}
\caption{Comparison of the predicted behavior of the order parameter versus the observed one for 
the different approximations (columns) and 
different networks (rows). If the behavior is correctly predicted by a given approximation for a certain network, the corresponding
entry contains the range of $k/k_c$ after $k/k_c = 1$ for which the approximation works well. A ``$+$'' indicates that the agreement 
possibly persists for larger values of $k$.
When ``NG'' appears in the corresponding entry in table \ref{transition}, no comparison is attempted and a ``-'' is entered. A ``-''
is entered when the perturbation theory is inapplicable ($\gamma < 5$), see Sec.~\ref{mftsec}. }
\label{orderparameter}
\end{table}

\section{Linear stability approach}\label{linear}

Partly as a precursor to the next section (Sec.~\ref{fluctuations}), in this section 
we discuss another approach that has the advantage of providing information on the dynamics of the system. 
We study the linear stability of the incoherent state by a method similar to that used in Ref.~\cite{daido}.
We assume that in the incoherent state 
the solution to Eq.~(\ref{eq:coupled}) is given approximately by
\begin{equation}\label{incoherent}
\theta_{n}^{0} = \omega_{n} t + \phi_{n},
\end{equation}
where $\phi_{n}$ is a random initial condition.
We introduce infinitesimal perturbations to this state by
\begin{equation}
\theta_{n} = \theta_{n}^{0} + \delta_{n}.
\end{equation}
In Appendix \ref{appendixb}, we assume that the perturbations grow as a function 
of time as $e^{s t}$, and obtain the eigenvalue equation
\begin{equation}\label{eq:eigequation0}
b_{n} = \frac{k}{2} \sum_{m=1}^{N} \frac{A_{nm} b_{m}}{s - i \omega_{m}}.
\end{equation}
We look for solutions $b_{n}$ of this equation that are independent of the frequencies
$\omega_{n}$ (similar to Assumption~$\bigstar$). 
Under this assumption, replacing $(s - i\omega_n)^{-1}$ in Eq.~(\ref{eq:eigequation0}) with its expected value, 
we get 
\begin{equation}\label{eq:eigequation2}
b_{n} = \frac{k}{2} \left<\frac{1}{s - i \omega}\right> \sum_{m=1}^{N} A_{nm} b_{m},
\end{equation}
where
\begin{equation}
\left<\frac{1}{s - i \omega}\right> = \int_{-\infty}^{\infty} \frac{g(\omega)d\omega}{s - i \omega}
\end{equation}
and the integration contour is defined in the causal sense [i.e., for $Re(s) > 0$ it is along the real axis, and for $Re(s) \leq 0$
it passes above the pole $\omega = -i s$].
We thus obtain the dispersion relation 
\begin{equation}\label{eq:dispersion}
1 = \frac{k\lambda}{2} \int \frac{g(\omega)d\omega}{s - i\omega},
\end{equation}
where, as in Sec.~\ref{self}, $\lambda$ is the largest eigenvalue of the adjacency matrix $A$.
Except for the presence of the eigenvalue $\lambda$, this is the known dispersion relation for the stability of the incoherent state
of the Kuramoto model \cite{strogatz}.
Under our assumption that $g(\omega)$ is even and decreases monotonically away from $0$ (Sec.~\ref{self}),
an unstable [$Re(s)> 0$] solution of Eq.~(\ref{eq:dispersion}) is real \cite{strogatz2} (note that, since $A$ is symmetric, 
$\lambda$ is real).
In order to find the critical coupling, we let $s \to 0^+$, $(s- i\omega)^{-1} \to i P(1/\omega) + \pi \delta(\omega)$.
Since $g(\omega)$ is symmetric, $\left<(s - i \omega)^{-1}\right> \to \pi g(0)$. 
According to Eq.~(\ref{eq:eigequation2}), the critical coupling
is then given by 
\begin{equation}\label{eq:kc2}
k_{c} = \frac{k_{0}}{\lambda},
\end{equation}
in agreement with the nonlinear approach. [We note, however, that, if $g(\omega)$ has multiple maxima, 
then the first instability can occur at $Im(s) \neq 0$ at a value of $k$ below that given by Eq.~(\ref{eq:kc2}).
This is why we have assumed that $g(\omega)$ decreases monotonically away from $\omega = 0$.]

\section{Effect of fluctuations}\label{fluctuations}

So far we have neglected the effect of the small fluctuations due to the finite number of connections per node. 
In our examples, we have presented networks that do not have nodes with small degree. However, in many networks
there is a large fraction of the nodes with small degree; in all our examples in Sec.~\ref{examples} there were no nodes 
with degree less than $50$ [$p(d) = 0$ for $d < 50$]. For example, scale free networks generated using the 
Barab\'{a}si-Albert method \cite{barabasi1} sometimes have parameters so that $\langle d \rangle = 6$.

In developing our theory, we neglected the time variations in Eq.~(\ref{eq:coupled2}), and worked thereafter with the 
average value of the phase of the locked oscillators. In order to gain insight into the effect of these fluctuations,
we will treat the time fluctuations as a noise term. 

The theory we present is heuristic and may be thought of as an expansion giving a small lowest order correction 
to the linear stability approach of Sec.~\ref{linear}
for large but finite $\langle d\rangle$. On the other hand, later in this section, we will apply this theory to numerical
examples where the finite size effect is not small, and we will find that the theory is still useful in that it correctly
indicates the trend of the numerical observations.

Like in Sec.~\ref{linear}, we consider perturbations to the incoherent state described by Eq.~(\ref{incoherent}).
As an approximation, we regard the coupling term in 
Eq.~(\ref{eq:coupled}), $f_n(t) \equiv k \sum_{m=1}^{N} A_{nm}\sin(\theta_{m} - \theta_{n})$, as a noise term. 
In addition to growing linearly with time, the phase of the oscillator $n$ will diffuse under the influence of this noise. 
We assume that
$\theta_n(t) = \phi_{n} + \omega_n t + W_n(t) $, where $W_n(t)$ is a random walk such that $\langle W_n(t)\rangle = 0$ and 
$\langle W_m(t)W_n(t)\rangle = 2 D_{nm} t$, and $\phi_{n}$ is an initial condition, which we assume to be randomly drawn
from $[0,2\pi)$. (In this section, by $\langle\dots\rangle$ we mean an expected value, i.e. and ensemble average, rather than
an average over $t$ or $n$.)

By using the linear approach of Section~\ref{linear}, the diffusion coefficients $D_{nm}$
will give us information on how the critical coupling strength differs from Eq.~(\ref{eq:kc2}).
The diffusion coefficients $D_{nm}$ are given by

\begin{equation}\label{eq:cruza}
D_{nm} = \int_0^{\infty}\langle f_n(t + \tau/2) f_m(t -\tau/2)\rangle d\tau
\end{equation}
\begin{equation}
= \int_0^{\infty} \sum_{j,k}A_{nj}\langle \sin(\theta_j^+ - \theta_n^+)A_{mk} \sin(\theta_k^- - \theta_m^-) \rangle d\tau,\nonumber
\end{equation}
where $+$ (respectively $-$) indicates evaluation at $t + \tau/2$ (respectively $t - \tau/2$).
Consider first the case $n \neq m$. The contribution of the terms with $\{j,n\}\neq \{k,m\}$ vanishes 
after the integration, and we obtain, using the symmetry of $A$,
\begin{equation}
D_{nm} = \frac{k^2}{2} A^2_{nm} \langle \sin(\theta_m^+ - \theta_n^+)\sin(\theta_n^- - \theta_m^-) \rangle.
\end{equation}
We now introduce our aforementioned assumption that $\theta_n(t) - \omega_n t$ is a random walk plus a random initial condition, 
$\theta_n(t) = \phi_{n} + \omega_n t  + W_n(t)$.
Using the identity $\sin(x)\sin(y) = (\cos(x-y) - \cos(x+y))/2$ and averaging over the initial phases $\phi_{n}$ we get
\begin{eqnarray}
D_{nm} = -\frac{k^2}{2}\int_0^{\infty}A^2_{nm} \langle \cos( \Delta W_m - \Delta W_n +\omega_{mn}\tau  ) \rangle d\tau,\hspace{3mm}
\end{eqnarray}
where $\Delta W_n \equiv W_n^+ - W_n^-$ and $\omega_{mn} \equiv \omega_m - \omega_n$.
We now use the fact that for a Gaussian random variable $x$ with variance $\sigma^2$ we have  
$\langle \cos(x)\rangle = Re\langle e^{ix}\rangle = Re( e^{i\langle x\rangle - \sigma^2/2})$. In our case,
$\langle x\rangle = \omega_{mn}\tau$ and 
$\sigma^2 = \langle (\Delta W_m - \Delta W_n)^2 \rangle = 2(D_n + D_m -2 D_{nm})\tau$,
where $D_n \equiv D_{nn}$.
After using this to compute the 
expected value, and performing the integration, we obtain for $n\neq m$
\begin{equation}\label{eq:des1}
D_{nm} = -\frac{k^2}{2} A^2_{nm} \frac{D_n + D_m -2 D_{nm}}{(D_n + D_m -2 D_{nm})^2 + \omega_{mn}^2}. 
\end{equation}
If $n = m$, the calculation proceeds along the same lines, but the nonvanishing terms in Eq.~(\ref{eq:cruza}) are those
for which $k = j$. Together with Eq.~(\ref{eq:des1}), this results in
\begin{equation}\label{eq:des2}
D_n = -\sum_{m \neq n} D_{nm}.
\end{equation} 

In principle, Eqs.~(\ref{eq:des1}) and (\ref{eq:des2}) can be solved for $D_n$ as a function of $k$ 
if the frequencies and the adjacency matrix are known.

In order to relate the diffusion coefficients to the critical coupling constant, we resort to the linear analysis of Sec.~\ref{linear}.
When noise is introduced in the linear approach,
Equation~(\ref{eq:eigequation0}) for the growth rate $s$ generalizes, as shown at the end of Appendix \ref{appendixb}, to 
\begin{equation}\label{eq:eigeD0}
b_{n} = \frac{k}{2} \sum_{m=1}^{N} \frac{A_{nm} b_{m}}{s + D_m - i \omega_{m}}.
\end{equation}
Since $Re(s) > 0$ corresponds to instability of the incoherent state, it is expected that 
the effect of the noise as reflected by 
positive $D_m$ is to shift the transition point so that the critical coupling constant is larger.

In order to solve for the growth rate $s$ for a given value of $k$, we rewrite Eq.~(\ref{eq:eigeD0}) as 
\begin{equation}\label{eq:eigeD1}
b = \frac{k}{2} D(s) A b,
\end{equation}
where $b$ is the vector with components $\{b_n\}$, $D(s)$ is the diagonal 
matrix $D(s) \equiv diag\{(s + D_m - i \omega_m)^{-1}\}$, and $A$ is the adjacency matrix.
The characteristic equation is
\begin{equation}\label{eq:eigeD2}
\det\left(\frac{k}{2} D(s) A - I\right) = 0,
\end{equation}
where $I$ is the $N\times N$ identity matrix. This implies
\begin{equation}\label{eq:eigeD3}
\det\left(\frac{k}{2} A - D(s)^{-1}\right) = 0,
\end{equation}
or 
\begin{equation}\label{eq:eigeD4}
\det\left(\frac{k}{2} A - diag\{D_m - i \omega_m\} - s I\right) = 0,
\end{equation}
that is, the growth rate $s$ is an eigenvalue of the matrix $M(k) \equiv (k/2) A - diag\{D_m - i \omega_m\}$.

For a given value of $k$, Eqs.~(\ref{eq:des1}) and (\ref{eq:des2}) can be solved iteratively. We have found that, by starting
from an initial guess for the values of $D_{nm}$ and repeatedly evaluating the right hand side of Eq.~(\ref{eq:des1}) in 
order to get the next approximation to the values of $D_{nm}$, convergence is achieved to a 
solution that is independent of the initial guess if the condition $D_{n} > 0$ is imposed. 
When the values of $D_{nm}$ have been found
for a given value of $k$, the relevant growth rate is calculated as the largest real part of the 
eigenvalues of the matrix $M(k)$ defined above.

As an example, we consider three networks with the degree of all nodes $d$ given by 
$d = 100$ in the first, $d = 50$ in the second and $d =20$ in the third one. 
In order to solve numerically the coupled equations, we work with a small number of nodes, $N = 500$. 
\begin{figure}[h]
\begin{center}
\epsfig{file = 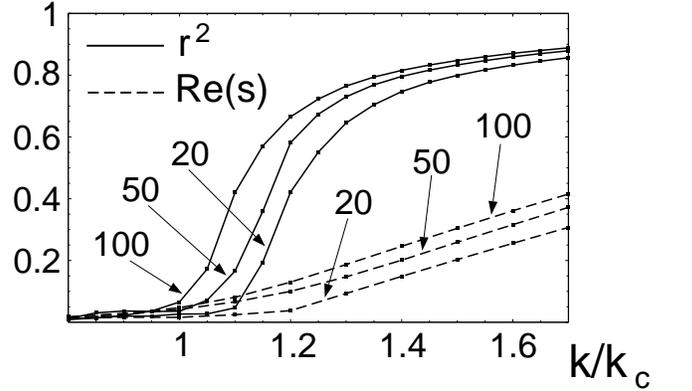, clip =  ,width=1.0\linewidth}
\caption{
Order parameter $r^2$ obtained from numerical solution of Eq.~(\ref{eq:coupled}) (solid lines) and growth rate $Re(s)$ 
(dashed lines) for a network with the degree of all nodes  $d = 20$, $d = 50$ and 
$d = 100$ as a function of $k/k_c$. The arrows indicate which network corresponds to the given curve.
} 
\label{figflu}
\end{center}
\end{figure}
In Fig.~\ref{figflu} we show the results for a realization of the three networks. 
The order parameter $r^2$ obtained from numerical solution of Eq.~(\ref{eq:coupled}) 
(solid lines) and the growth rate obtained from Eqs.~(\ref{eq:des1}), (\ref{eq:des2}) and (\ref{eq:eigeD4}) 
(dashed lines) are plotted as a function of $k/k_c$. 
The arrows indicate which network corresponds to the given curve.
We observe that, as the connections per node are decreased, the transition point shifts to larger values of the coupling constant. 
This trend is reproduced by the growth rate curves, which are displaced to the right for smaller values of the degree.

We emphasize that the theory we described above is applicable to
networks for which $\langle d \rangle$ is large but finite. However, in Fig.~\ref{figflu} we applied the theory to cases in 
which $\langle d \rangle$ is not very large.
Although we do not expect the theory to be valid in this case, we find that it correctly describes the trend present in the numerical 
observations, i.e., a shifting of the transition to coherence to larger values of the critical coupling 
as nodes of small degree become important.

\section{Discussion}\label{discussion}

A transition to coherence in large networks of coupled oscillators 
should be expected at a critical value of the coupling strength
which is determined by the largest eigenvalue of the adjacency matrix of the network and its associated eigenvector. 
In the all-to-all case, the largest eigenvalue is $N -1 \approx N$ and thus the Kuramoto result $k_c = k_0/N$ is recovered. 
The largest eigenvalue of the adjacency matrix of a network is of both theoretical and practical importance, and thus
its properties have been studied in some detail \cite{chung, goh}. We remark that our analysis allows the case of nonuniform 
interaction strengths by introducing continuous values in the entries of the adjacency matrix $A$. 

We developed different approximations in order to describe the transition to coherence in terms of an appropriately defined
order parameter which generalizes the parameter used in the classical Kuramoto model \cite{ichinomiya}. 
See Table~\ref{abbreviations} and Figure~\ref{fig:diagram} for a summary of the approximations and assumptions.
The time averaged theory (TAT) provided the most accurate description of the behavior of the order parameter, and assumes knowledge of
the adjacency matrix $A_{nm}$ and the individual frequencies $\omega_n$. 
The frequency distribution approximation (FDA) also provides a good approximation 
but does not require knowledge of the individual frequencies. These approximations yield equations 
that have to be solved numerically. The time required to numerically solve these equations is, however, much less than that 
required to numerically integrate the original differential equations.
The perturbation theory (PT) yields analytic expressions for the order parameter when close to the transition in terms of the 
largest eigenvalue of the adjacency matrix and its associated eigenvector, but is limited to 
networks with a relatively homogeneous degree distribution. 
The mean field theory (MF) \cite{ichinomiya} is obtained 
by introducing the additional assumption that
the components of the eigenvector associated with the largest eigenvalue are proportional to the degree of the corresponding node.
This does not necessarily have to be the case when close to the transition, and 
because of this extra assumption, we expect the other approximations to more generally accurately describe the 
transition than the mean field theory.
Figs.~\ref{figama}(a) and (b) show that for the particular case of scale free networks with $N = 2000$, 
$\gamma = 2$ and $\gamma = 2.5$ this is the case. In general,
we observed that for low values of the exponent $\gamma$ (see Fig.~\ref{figama}) the mean field approximation and the 
perturbative approximation yield
different critical coupling strengths. 
The mean field theory has the advantage that analytic expressions can be computed without the need of solving 
the eigenvalue problem for the adjacency matrix, and could
be useful when only limited information  is available about the
network. However, in general, our results suggest that 
one of the other approximations mentioned above should be used.

We remark that even though the time averaged theory, the frequency distribution approximation and the perturbation theory
require in principle knowledge of the full matrix $A$, knowledge of the degree distribution may be enough in some cases. 
As in our examples, an adjacency matrix $A$ can be generated randomly with a given degree distribution. Our results
indicate that even this limited reconstruction of the original network might improve the mean field results 
(see Sec.~\ref{examples}).

Our assumptions restrict the class of networks for which the results apply. 
We assumed that sufficiently near the onset of synchronization 
each node is coupled to many {\it locked} oscillators. In practice this implies that most nodes should have a high degree. 
This is an important restriction for our theory. In Sec.~\ref{examples} we used networks with a minimum degree of $50$.  
As mentioned before, we observed that in networks with small average degree (about $20$), the observed critical coupling was larger 
than the one predicted by our theory. By including the previously neglected time fluctuations, we 
developed a heuristic theory in Sec.~\ref{fluctuations} which correctly predicts the trend observed 
in the numerical simulations. As the nodes with small
degree become important, both our theory and the numerical observations indicate that the transition to synchrony occurs 
at larger values of the coupling strength.

In conclusion, we have developed a theory predicting the critical coupling for the transition from incoherence to coherence in 
large networks of coupled oscillators. We found that for a large class of networks, a transition to coherence should be 
expected at a critical value of the coupling strength which is determined by the 
largest eigenvalue of the adjacency matrix of the network. 
We developed and compared various approximations to the order parameter past the transition, and studied the effect of the 
fluctuations caused by finite size effects.

This work was sponsored by ONR (Physics) and by
NSF (contracts PHYS 0098632 and DMS 0104087).

\appendix

\section{}\label{appendixa}

In this Appendix we show that, using Assumption~$\bigstar$, we can neglect the sum over the unlocked 
oscillators in Eq.~(\ref{eq:beta}),
\begin{equation}\label{eq:sum2}
\sum_{\left|\omega_{m}\right| > k r_m}^N A_{nm}\langle e^{i\theta_m}\rangle_t.
\end{equation}
We will follow to some extent Chapter $12$ of Ref.~\cite{pikovsky}. The time average is given by 
\begin{equation}
\langle e^{i\theta_m}\rangle_t = \int_{-\pi}^{\pi} e^{i\theta} p_m(\theta) d\theta.
\end{equation}
where $p_m(\theta)d\theta$ is, given the connections of node $m$ 
and its natural frequency $\omega_m$, the probability that its phase $\theta_m$ lies in the interval
$[\theta,\theta + d\theta)$.
It satisfies $p_m(\theta)\propto 1/\left|\dot{\theta}\right|$.
Including the normalization we have, 
neglecting the term $h_m$ in Eq.~(\ref{eq:coupled2}), 
\begin{equation}
p_m(\theta) = \frac{\sqrt{\omega_m^2 - k^2 r_m^2 }}{2\pi\left|\omega_m - k r_m \sin(\theta - \psi_m)\right|}.
\end{equation}
The sum in Eq.~(\ref{eq:sum2}) can be written as
\begin{widetext}
\begin{equation}\label{eq:sum3}
\sum_{\left|\omega_{m}\right| > k r_m}^N A_{nm}\langle e^{i\theta_m}\rangle_t =
\sum_{\left|\omega_{m}\right| > k r_m}^N A_{nm}\sqrt{\omega_m^2 -k r_m^2} 
sign(\omega_m) \frac{1}{2\pi}\int_{-\pi}^{\pi} \frac{e^{i\theta}(\omega_m + 
k r_m \sin(\theta - \psi_m))d\theta}{\omega_m^2 - k^2 r_m^2 \sin^2(\theta -\psi_m)}. 
\end{equation}
\end{widetext}
The integral of the first term vanishes since 
the $2\pi$-periodic integrand changes sign under the transformation $\theta \to \theta +\pi$.
We are left with 
\begin{widetext}
\begin{equation}\label{eq:sum4}
\sum_{\left|\omega_{m}\right| > k r_m}^N A_{nm}\langle e^{i\theta_m}\rangle_t =
\sum_{\left|\omega_{m}\right| > k r_m}^N A_{nm}\sqrt{\omega_m^2 -k r_m^2} k r_m
sign(\omega_m) \frac{1}{2\pi}\int_{-\pi}^{\pi} \frac{e^{i\theta}\sin(\theta - \psi_m)d\theta}{\omega_m^2 - k^2 r_m^2 \sin^2(\theta -\psi_m)}. 
\end{equation}
\end{widetext}
In this sum, $sign(\omega_m)$ is independent of $\omega_m^2$ and, using Assumption~$\bigstar$, it is independent of
$r_n$ and $\psi_n$ as well. If there are many terms in the sum, it will be then of order $\sqrt{d_n}$ due to 
the symmetry of the frequency distribution, and thus will be small compared with the sum over the locked oscillators,
which is of order $d_n$ [see Eq.~(\ref{eq:betass})]. Note that we did not use here the full strength of 
Assumption~$\bigstar$, since we only required the sign of $\omega_m$ to be independent of $r_m$ and $\psi_m$.

\section{}\label{appendixa2}

Here we show that for sufficiently large $N$ and a power law degree distribution of the degrees, 
$p(d)\propto d^{-\gamma}$, the mean field approximation $\langle d^2 \rangle/\langle d \rangle$
underestimates $\lambda$ for $\gamma > 3$.
We base our argument in the results of Ref.~\cite{chung}: if for a random graph 
$\sqrt{d_{max}} > \langle d^2 \rangle/\langle d \rangle (\log N)^2$,
then $\lambda \sim \sqrt{d_{max}}$ almost surely as $N\to \infty$, where $d_{max}$
is the largest expected degree.

In the case under consideration ($\gamma > 3$), $\langle d^2 \rangle/\langle d \rangle$ converges
to the finite value $\langle d^2 \rangle_{\infty}/\langle d \rangle_{\infty}$
[$\langle\dots\rangle_{\infty}$ is defined by Eq.~(\ref{x})], while $d_{max}$
diverges as $N^{1/(\gamma -1)}$ \cite{newman1}. Thus, for large enough $N$, the conditions for $\lambda \sim \sqrt{d_{max}}$
will be satisfied, since $N^{1/(\gamma -1)}/(\log N)^4 \to \infty$ as $N \to \infty$. 
While $\lambda \sim \sqrt{d_{max}} \to \infty$ as $N\to \infty$, the mean field approximation 
$\langle d^2 \rangle/\langle d \rangle$ remains finite.

We can estimate an upper bound on how large $N$ needs to be for this discrepancy to be observed. For large $N$,
$\langle d^2 \rangle/\langle d \rangle \sim d_0$, where $d_0$ is the minimum degree [$p(d) = 0$ for $d < d_0$].
The maximum degree is approximately given by $d_{max} \sim d_0 N^{1/(\gamma - 1)}$ \cite{newman1}. Inserting these estimates
in the condition $\sqrt{d_{max}} \sim \langle d^2 \rangle/\langle d \rangle (\log N)^2$ we obtain
\begin{equation}
N \sim d_0^{\gamma-1} (\log N)^{4(\gamma -1)}.
\end{equation}
As an example, for $\gamma = 4$ and $d_0 = 20$, the upper bound is approximately $N \sim 10^{25}$,
a far larger system than we can simulate.

\section{}\label{appendixb}

In this Appendix we study the linear stability of the incoherent state by a method similar to that presented in Ref.~\cite{daido}. 
As described in Section \ref{linear}, we assume that in the incoherent state 
the solution to Eq.~(\ref{eq:coupled}) is given approximately by
\begin{equation}
\theta_{n}^{0}(t) \approx \omega_{n} t + \phi_{n},
\end{equation}
where $\phi_{n}$ is an initial condition.
We introduce infinitesimal perturbations to this state by
\begin{equation}
\theta_{n} = \theta_{n}^{0} + \delta_{n}.
\end{equation}
Linearizing Eq.~(\ref{eq:coupled}), we get 
\begin{equation}\label{eq:linear}
\dot{\delta}_{n} = k \sum_{m=1}^{N} A_{nm} \cos(\theta_{m}^0 - \theta_{n}^0)\delta_{m} +  \mu_{n} - \nu_{n}\delta_{n},
\end{equation}
where $\mu_n = k \sum_{m=1}^{N} A_{nm} \sin(\theta_{m}^0 - \theta_{n}^0)$ and 
$\nu_n = k \sum_{m=1}^{N} A_{nm} \cos(\theta_{m}^0 - \theta_{n}^0)$. 
As before, we assume that the number of links to node $n$ is so large that, due to the incoherence, we may neglect the 
terms $\mu_n$ and $\nu_n$.
With this simplification, Eq.~(\ref{eq:linear}) can be recast as an integral equation as follows:
\begin{widetext}
\begin{eqnarray}
\delta_{n}(t) = k \int_{-\infty}^{t}dt'
\sum_{m=1}^{N} A_{nm} \delta_{m}(t') \cos[\theta_{m}^0(t') - \theta_{n}^0(t')]\\
= \frac{k}{2} \int_{-\infty}^{t}dt'
e^{-i\theta_n^0(t')} \left( \sum_{m=1}^{N} A_{nm}e^{i\theta_{m}^0(t')}\delta_{m}(t') + 
 \sum_{m=1}^{N} A_{nm} e^{i[2\theta_{n}^0(t') - \theta_{m}^0(t')]} \delta_{m}(t')\right).\nonumber
\end{eqnarray}
\end{widetext}
Multiplying by $A_{jn} e^{i\theta_{n}^{0}(t)}$, summing over $n$ and defining 
$B_{n}(t) \equiv \sum_{m=1}^{N} A_{nm} \delta_{m}(t) e^{i\theta_{m}^{0}(t)}$,
we get 
\begin{widetext}
\begin{equation}\label{eq:messy}
B_{j}(t) = \frac{k}{2}\int_{-\infty}^{t}dt'\sum_{n=1}^{N} A_{jn} e^{i[\theta_{n}^0(t) - \theta_{n}^0(t')]}
\left(B_{n}(t') + e^{2i  \theta_n^0(t')}B_{n}^*(t') \right).
\end{equation}
\end{widetext}
We assume that the quantities $B_{n}$
grow exponentially with time as $B_{n}(t) = b_{n} e^{s t}$, where $Re(s) > 0$. Inserting this ansatz in Eq.~(\ref{eq:messy}), 
and performing the integration we get
\begin{equation}\label{eq:eigextra}
b_{j} = \frac{k}{2} \sum_{n=1}^{N} \frac{A_{jn} b_{n}}{s - i \omega_{n}} 
+ \frac{k}{2}e^{2 i Im(s) t} \sum_{n=1}^{N} \frac{A_{jn} b^*_{n} e^{2 i \theta_n^0(t)}}{s^* + i \omega_{n}}.
\end{equation}
The second sum is very small due to the incoherence of the $\theta_n^0$'s. So, changing indices, we are left with the 
eigenvalue equation
\begin{equation}\label{eq:eigequation}
b_{n} = \frac{k}{2} \sum_{m=1}^{N} \frac{A_{nm} b_{m}}{s - i \omega_{m}},
\end{equation}
as claimed in Section \ref{linear}.

If, as proposed in Section \ref{fluctuations}, there are fluctuations in the values of $\theta_n^0(t)$ such that 
$\theta_n^0(t) = \omega_n t + \phi_{n} + W_n(t)$, where $W_n(t)$ is a random walk such that $\langle W_n(t)\rangle = 0$
and $\langle W_n(t)^2\rangle = 2 D_n t$, we take the expected value of Eq.~(\ref{eq:messy}). 
We use the fact that for a Gaussian random variable $x$ with variance $\sigma^2$ we have 
$\langle e^{ix}\rangle =  e^{i\langle x\rangle - \sigma^2/2}$. In this case, $x = \omega_m(t' -t)$ and 
$\sigma^2 = 2 D_m (t -t') $. We obtain after performing the integration
\begin{equation}\label{eq:eigeD8}
b_{n} = \frac{k}{2} \sum_{m=1}^{N} \frac{A_{mn} b_{m}}{s + D_m - i \omega_{m}}.
\end{equation}

\end{document}